\documentclass{IEEEtran}
\usepackage{graphicx,cite}
\usepackage{epsfig}
\usepackage[cmex10]{amsmath}
\usepackage{array}
\usepackage{fixltx2e}
\usepackage{amsfonts}

\begin{document}
\title{Adaptive Differential Feedback in Time-Varying Multiuser MIMO Channels}

\author{\IEEEauthorblockN{Muhammad Nazmul Islam\IEEEauthorrefmark{0}
 and Raviraj Adve\IEEEauthorrefmark{1} \\
 }
 \IEEEauthorblockA{\IEEEauthorrefmark{0}Dept.~of Elec.~and Comp.~Eng.,
WINLAB, Rutgers University, Email: mnislam@winlab.rutgers.edu \\
\IEEEauthorrefmark{1}Dept.~of Elec.~and Comp.~Eng., 
University of Totonto, Email: rsadve@comm.utoronto.ca}}


\maketitle

\begin{abstract}

In the context of a time-varying multiuser
multiple-input-multiple-output (MIMO) system, we design 
recursive least squares based adaptive predictors and
differential quantizers to minimize the sum mean squared error of the 
overall system. Using the fact that the scalar entries of the
left singular matrix of a Gaussian MIMO channel becomes ``almost''
Gaussian distributed even for a small number of transmit antennas, we
perform adaptive differential quantization of the relevant
singular matrix entries. Compared to the algorithms in the existing differential feedback
literature, our proposed quantizer provides three advantages: first,
the controller parameters are flexible enough to adapt themselves to
different vehicle speeds; second, the model is backward adaptive i.e.,
the base station and receiver can agree upon the predictor and variance
estimator coefficients without explicit exchange of the parameters;
third, it can accurately model the system even when the correlation
between two successive channel samples becomes as low as 0.05. Our
simulation results show that our proposed method can reduce the
required feedback by several kilobits per second for vehicle speeds up to 20 km/h
(channel tracker) and 10 km/h (singular vector tracker). 
The proposed system also
outperforms a fixed quantizer, with same feedback overhead, in terms of
bit error rate up to 30 km/h.

\end{abstract}


\section{Introduction}

The advantages of multiuser multiple-input-multiple-output (MIMO)
has led to its inclusion in the standard proposals for fourth generation wireless
systems, e.g., mobile WiMax~\cite{Wimax}. The best MIMO system
performance can be achieved when channel state information (CSI) is
available at the transmitter~\cite{Palomar}. In a frequency division
duplexing system, downlink CSI needs to be estimated at the receiver,
quantized and provided to the base station via an uplink feedback
channel. Recent work suggests that this might also be required in
broadband time division duplex systems~\cite{Haartsen}. Therefore,
reducing feedback overhead while providing accurate CSI plays an
important role in multiuser transceiver design. In the available literature, scalar quantization
(SQ)~\cite{Rao,Adam:a}, vector quantization (VQ)~\cite{Jindal:a} and
matrix quantization~\cite{Jindal:b} have all been used to quantize CSI.
SQ has been included in several standards (e.g.IEEE 802.11n~\cite{Heath:a})
due to its linear computational complexity. This paper focuses on
the reduction of feedback overhead in time varying channels that employs SQ.
We assume perfect channel estimation and delay-free noiseless feedback and focus
on quantization only.

The feedback overhead can be
significantly reduced using differential feedback by exploiting
temporal correlation of the channel~\cite{Love:b,Heath:b,Love:c}. Most
of these works model the channel as a first order Gauss-Markov process
and use a \emph{fixed differential quantizer}.
The authors assume that the transmitter and receiver agree on the value
of the parameters in the Markov chain. Since this assumption does not
hold in non-stationary channels, there has been some research in
adaptive delta-modulation (ADM) based feedback~\cite{Rao,Adam:a}. Both
these works quantize the difference between the previous
and current samples with a one-bit quantizer. However, the
authors give suitable step size controller parameters for
pedestrian velocities (up to 4 km/h) only.

The lack of flexibility of the available differential feedback methods
motivates us to investigate adaptive differential feedback in
time-varying multiuser channels. This paper makes the following two
contributions: First, based on the linear least square (LLS) based 
adaptive differential speech quantizer model proposed by Stroh~\cite{Stroh}, we develop a 2-bit
recursive least square (RLS) adaptive differential feedback in a time-varying environment.
Second, we design RLS adaptive tracking of the
singular vector entries of each users' channel matrix and show
that, if the number of data streams is less than the total number of
receive antennas, this method reduces feedback overhead. Both these
methods can lead to reducing the required feedback overhead by several kBits/sec
in modern wireless communication standards.

\emph{Notation}: Lower case, e.g., $ n $ or $L_k$, denote scalars while lower
case bold face, e.g., $ \mathbf{h} $ means a column vector. Upper case
boldface, e.g., $\mathbf{V} $ denotes a matrix. The superscripts
$(\cdot)^T $ and $ (\cdot)^H $ denote the transpose and conjugate
transpose operators respectively. tr [$\cdot$] indicates the trace
operator. 
$ E [\cdot] $ denotes the expectation operator.
$\mathbf{I} $ is reserved for the identity matrix. $ diag(\mathbf{x}) $
denotes a diagonal matrix with non-zero entries taken from
$\mathbf{x}$. $\mathbf{A} (:,1:L)$ denotes the leftmost L
columns of $\mathbf{A}$.

This paper is organized as follows: Section~\ref{sec:sysmodel} develops
the system model while Sections~\ref{sec:DiffQuantChannel}
and~\ref{sec:DiffQuantSingVec} develop adaptive differential
quantization for the channel entries and singular vectors respectively.
Section~\ref{sec:NumericalResults} presents the simulation results.
Finally, Section~\ref{sec:Conclusions} wraps up the paper.

\section{System Model}
\label{sec:sysmodel}

We consider linear precoding with quantized channel knowledge in a
multiuser MIMO environment where each users' vehicle is moving at an unknown speed.
Specifically, we consider a single base
station equipped with $ M $ transmit antennas and $K$ independent
users. User $k$ has $N_k$ antennas and receives $L_k$ data streams. Let
$L = \sum_k L_k$ and $N = \sum_k N_k$. 
Let $\mathbf{U} \in \mathcal{C}^{M \times L}$ and $\mathbf{p} \in \mathcal{R}^L$
represent the beamformer matrix and power allocation vector respectively. $\mathbf{P} = diag(\mathbf{p})$.
$tr[\mathbf{P}] \leq P_{max}$ where $P_{max}$ is the total transmission power.
The overall data vector is 
$ \mathbf{x} = \left[x_1, x_2, \dots, x_L \right]$. 
The $N_k \times M$ block fading channel, $\mathbf{H}_k^H$, between the base
station (BS) and the user is assumed to be flat. The singular value
decomposition of $\mathbf{H}_k$ is given by $ \mathbf{H}_k =
\mathbf{A}_k \mathbf{\Sigma}_k \mathbf{B}_k$.
 $\mathbf{\Sigma}_k \in \mathcal{R}^{M \times N_k}$ contains the
singular values. 
global channel matrix is $ \mathbf{H}^H $, with $ \mathbf{H} = \left [
\mathbf{H}_1,...,\mathbf{H}_k \right] $.

In the downlink, user $ k $ receives
\begin{equation}
\mathbf{y}^{DL}_k = \mathbf{H}^H_k \mathbf{U} \mathbf{\sqrt{P}} \mathbf{x} + \mathbf{n}_k,
\end{equation} %
where $ \mathbf{n}_k $ represents the additive white Gaussian noise at
the receiver with $ E\left[\mathbf{n}\mathbf{n}^H\right] = \sigma^2
\mathbf{I}_{N_k}$. 
To estimate its own transmitted symbols, from
$\mathbf{y}^{DL}_k $, user $k$ forms $\hat{\mathbf{x}}_k =
\mathbf{V}^H_k \mathbf{y}^{DL}_k$.	Let $ \mathbf{V} $ be the $N \times
L$ block diagonal global decoder matrix, $ \mathbf{V} = diag \left (
\mathbf{V}_1, ..., \mathbf{V}_K \right) $. Overall
\begin{equation}
\hat{\mathbf{x}} =
\mathbf{V}^H\mathbf{H}^H\mathbf{U}\sqrt{\mathbf{P}}\mathbf{x} +
\mathbf{V}^H\mathbf{n} = \mathbf{F}^H \mathbf{U} \sqrt{\mathbf{P}} \mathbf{x} +
\mathbf{V}^H \mathbf{n}
\end{equation}
%
We define the $M\times L$ matrix
$\mathbf{F} = \mathbf{HV} $ with $ \mathbf{F} = \left [\mathbf{F}_1,
\dots , \mathbf{F}_K \right]$.  $\mathbf{F}_k = \mathbf{H}_k \mathbf{V}_k$.
To ensure resolvability, $ L \leq M $ and $ L_k \leq N_k $.
%
%
%
%
%
\begin{figure}[t]
 \epsfig{figure=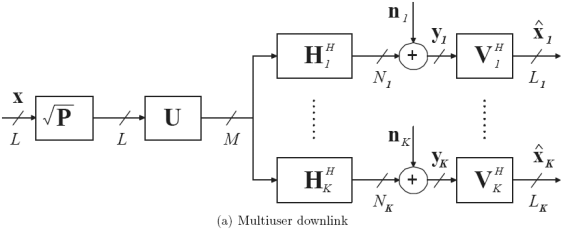,width=75mm}
 \caption{Block Diagram of Multiuser MIMO Downlink}  \label{fig:MUdownlink}
\end{figure} %
In \cite{Nazmul:c}, we showed that the sum mean squared error of the whole system can be
written as,
\begin{eqnarray}
SMSE & = & \sum_{i=1}^{L} E \left[||x_i - \hat{x}_i||^2 \right]    \\
& = & L - M + \left(\sigma^2 + \sigma^2_E P_{max} \right) tr \left[\mathbf{J}^{-1} \right]  \label{eq:SMSE}
\end{eqnarray}
$\mathbf{J} = \mathbf{F} \mathbf{Q} \mathbf{F}^H + \left(\sigma^2 + \sigma^2_E P_{max} \right) \mathbf{I}_M$.
$\sigma^2_E$ denotes the quantization error variance of the feedback model and
 $\mathbf{Q}$ represents the virtual uplink power allocation matrix.
\eqref{eq:SMSE} is a nonincreasing
function of $\sigma^2_E$. Our objective is to minimize $\sigma^2_E$ by
designing a backward adaptive differential feedback system.
%
\begin{table}
\caption{(Channel Parameters)} \label{tab:ChannelParam}
\begin{center}
\begin{tabular}{|l|l|} \hline
Parameter & Value(units) \\ \hline
Carrier Frequency ($\mathbf{f}_c$) & 2.5 GHz \\ \hline
Channel Sampling Rate ($\mathbf{f}_s$) & 200 Hz \\ \hline
Frame Duration ($\mathbf{T}_{f_r}$) & 5 ms \\ \hline
\end{tabular}
\end{center}
\end{table}

\textit{Channel Mobility:} The channels are temporally correlated and
assumed to follow a modified version of Jakes' model~\cite{Zheng}.
Here, channel parameters are selected (as in
Table~\ref{tab:ChannelParam}) to represent typical values for the WiMax
standard~\cite{Wimax}.

\indent \textit{Feedback Model:}  We use two
different feedback methods. In the first method, 
the full channel matrix is quantized and sent back to the base station
(BS). The receivers expend 2 bits on the differential quantization of the real and imaginary
parts of each scalar channel entry based on minimum Euclidean distance. The BS
uses the following channel model:
%
\begin{equation}
\hspace*{0.1in} \mathbf{H} = \mathbf{\widehat{H}} + \mathbf{\widetilde{H}}
\label{channel_model1}
\end{equation}
Here $\mathbf{\hat{H}}$ and $\mathbf{\widetilde{H}}$ denote the
quantized channel and error in channel feedback respectively. This channel 
model is used to find the optimal $\mathbf{F}$ through iteration~\cite{Khachan}.
Also, $\sigma^2_E = E \left[|h - \hat{h}|^2 \right]$,  
 $h$ and $\hat{h}$ denote the original and quantized channel entry.

In the second method, the receivers use, as $ \mathbf{V}_k $, the $L_k $ right singular
vectors corresponding to the maximum singular values of $ \mathbf{H}_k$. 
So, $ \mathbf{V}_k = \mathbf{B}_k(:,1:L_k) $ Therefore,
$\mathbf{F}_k = \mathbf{A}_{k}(:,1:L_k) \times \mathbf{\Sigma}_{k}(:,1:L_k)$. 
The receivers perform 2
bits adaptive differential quantization of each real and
imaginary scalar entry of $\mathbf{A}_{k}(:,1:L_k)$ and 2 bit fixed quantization 
of the entries of $\mathbf{\Sigma}_{k}(:,1:L_k)$.
The BS assumes the following model,
\begin{equation}
\hspace*{0.1in} \mathbf{F} = \mathbf{\widehat{F}} + \mathbf{\widetilde{F}}
\label{channel_model2}
\end{equation}
Here, $\mathbf{\widehat{F}}$ and $\mathbf{\widetilde{F}}$ represent the
quantized effective channel and error in the feedback respectively.
Here, $\sigma^2_E = E \left[|f - \hat{f} \right|^2]$. $f$ and $\hat{f}$ denote
the scalar entries of $\mathbf{\widehat{F}}$ and $\mathbf{\widetilde{F}}$ respectively.

The linear precoding algorithms of these two feedback models 
can be found in~\cite{Khachan} and ~\cite{Nazmul:b} respectively. We do not include those
here for brevity.

\section{Quantization of Channel Entries} \label{sec:DiffQuantChannel}
\begin{figure}[t]
 \epsfig{figure=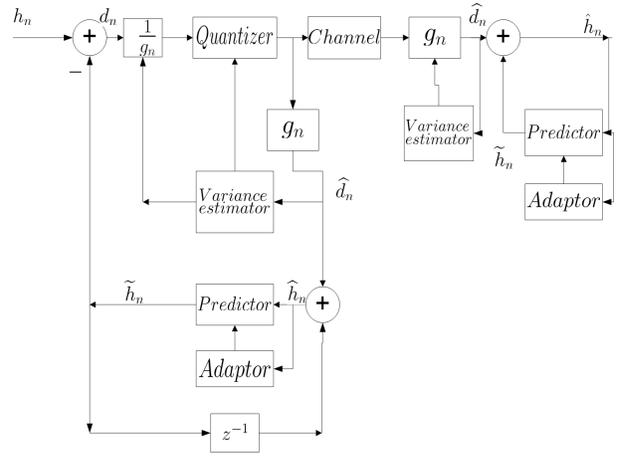,width=80mm, height = 60mm}
 \caption{Block Diagram of the adaptive differential quantizer}  \label{fig:ADPCM}
\end{figure} %
Stroh~\cite{Stroh} proposed the differential quantizer model, shown in
Fig.~\ref{fig:ADPCM}, for speech quantization. We use it to perform
adaptive differential channel quantization. The left and right sides
of the channel block are located at the receiver and base station
respectively. A unit variance Gaussian quantizer~\cite{Lloyd} is
used in the quantizer block. Let $h_n$ and $\hat{h}_n$ represent
the original and quantized channel parameter at the $n^\mathrm{th}$
instant. $d_n$ is the difference between $h_n$ and predicted channel
$\tilde{h}_n$. $g_n$ normalizes the variance of the difference signal,
i.e., avoids granular noise and overloading. Thus, $g_n$ enables 
the quantizer block to adapt to different speeds.
$\hat{d}_n = d_n + qn_n$ where $qn_n$ is the quantization noise.

Using the symmetry of the receiver and BS, $\hat{h}_n = h_n +
qn_n$~\cite{Gibson,Stroh}. Here, the adaptor block controls the
predictor coefficients and the variance estimator block estimates
$g_n$. The predictor coefficients and variance estimator parameters
depend on $\mathbf{\hat{h}}_n$ and $\mathbf{\hat{d}}_n$, rather than on
$\mathbf{h}_n$ and $\mathbf{d}_n$. Therefore, unlike the differential
feedback model proposed in~\cite{Heath:b,Love:b,Love:c}, the BS can
reproduce the predictor and variance estimator parameters without the
explicit transmission of the coefficients. Since the real and imaginary
part are quantized separately, $g_n = 2 \sigma^2_E$.

\subsection{Parameter Selection of the adaptive quantizer}
\subsubsection{LLS Based Predictor}
Stroh proposed the following LLS based predictor in the speech quantizer~\cite{Stroh},
\begin{equation}
\tilde{h}_n = \sum^{T}_{j=1} w_{j,n} \hat{h}_{n-j}
\end{equation}
Here, $T$ is the predictor order and $\mathbf{w}_{j,n}$ is the
$j^\mathrm{th}$ weight coefficient at the $n^\mathrm{th}$ time instant.
The predictor coefficients are computed to minimize the mean squared
error
\begin{equation}
\epsilon^2 = \frac{1}{Lp} \sum^{L_p}_{i=1} \left [\hat{h}_{n-i} -
            \sum^{T}_{j=1} w_{j,n} \hat{h}_{n-i-j} \right]^2 \label{eq:error_term}
\end{equation}
Here, $L_p$ is the learning period. $\epsilon^2$ is the fitting or
average prediction error.
The weights are calculated through Weiner filtering~\cite{Haykin}.  \newline
%
%
%
\begin{table}
\caption{(Design Parameters)}\label{tab:DesignParam}
\begin{center}
\begin{tabular}{|l|l|l|l|l|l|l|l|} \hline
Parameter & $\lambda$ & $k_2$ & $k_1$ & $T$ & $L_P$ \\ \hline
Value & 0.98 & 0.9 & 1.1 & 2 & 100 \\ \hline
\end{tabular}
\end{center}
\end{table}
\subsubsection{RLS Based Predictor} LLS predictors
perform close to ideal Weiner filter
predictors in terms of quantization error reduction. However, reducing the steady state error to acceptable
levels requires increasing the learning period and an attendant
increase of transient time. This motivates us to design
a recursive least square based backward predictor~\cite{Haykin}.
The weights $\mathbf{w} (n)$ are calculated by solving the Weiner-Hopf
equation, $\mathbf{\Phi}(n) \mathbf{w}(n) = \mathbf{\psi}(n)$; 
$\mathbf{w(n)}$ is the $T \times 1$ vector of predictor coefficients,
at the $n^\mathrm{th}$ time instant. $\mathbf{\Phi(n)} \in \mathcal{R}^{T \times T}$ and 
$\mathbf{\psi(n)} \in \mathcal{R}^{T \times 1}$ are given by,
\begin{eqnarray}
\mathbf{\Phi(n)} & = & \lambda \mathbf{\Phi(n-1)} + \mathbf{\hat{h}}(n-1) \mathbf{\hat{h}}^H(n-1)
                                                                            \label{eq:Phi_RLS2} \\
\mathbf{\psi(n)} & = & \lambda \mathbf{\psi(n-1)} + \mathbf{\hat{h}}(n-1) \hat{d}^H(n-1)
                                                                            \label{eq:psi_RLS2}  
\end{eqnarray}
Here, $\mathbf{\hat{h}}(i) = \left[\hat{h}_i, \cdots, \hat{h}_{i-T+1}
\right]$ and $\lambda$ is the memory factor. \newline

\subsubsection{RLS Variance Estimators}
%
%

For the RLS variance estimator, $g_n$ is calculated as~\cite{Haykin},
\begin{eqnarray}
v_n & = & \sum^{n-1}_{i=1} k_2^{n-1-i} \hat{d}^2_i \label{eq:var_RLS}  \\
g_n & = & k_1 \sqrt{(1-k_2) \left (k_2 v_{n-1} +  \hat{d}^2_{n-1} \right)}. \label{eq:var_RLS2}
\end{eqnarray}
%
%
Here, $k_1$ and $k_2$ denote the bias compensator and memory factor respectively in RLS variance estimator.
Table~\ref{tab:DesignParam} lists the parameters of the adaptive differential quantizer.
The parameters were chosen via numerical simulations.
\section{Quantization of singular vectors}\label{sec:DiffQuantSingVec}


The scalar entries of $\mathbf{A}_k(:,1:L_k)$, the left singular vector
matrix of $\mathbf{H}$, can be adaptively differentially quantized
using the same model shown in Fig.~\ref{fig:ADPCM}. Note that both the
adaptive predictors proposed in the previous section do not assume any
particular model of the signal; they try to find the ``best" predicted
value based on the past observations. However, 
the quantizer in the proposed adaptive differential feedback model
assumes a Gaussian distributed input. Therefore, if we can show the
entries of $\mathbf{A}_k$ to be approximately Gaussian, the model of
Fig.~\ref{fig:ADPCM} can be readily applied to track $\mathbf{A}_k$~\cite{Petz}.

The matrix of singular vectors of a rectangular Gaussian matrix is called a Haar
matrix~\cite{Petz}.
\newline \indent \emph{Lemma 1:} If $\mathbf{A}_k$ is a $\mathbb{C}^{M \times M}$ Haar matrix,
$ E \left[ |\mathbf{A}_{ij}|^2 \right] = \frac{1}{M}$ , $1 \leq i,j
\leq M$.
\newline
\indent \emph{Proof:} See ~\cite{Petz}.
\newline
\indent \emph{Lemma 2}: The probability distribution of $\sqrt{M}$
times the Haar matrix $\mathbf{A}_k$, approaches the standard complex
Gaussian measure as $M \rightarrow \infty$.
\newline
\indent \emph{Proof:} See 4.2.11 of ~\cite{Petz}.

In practice, the entries of the Haar matrix approach a Gaussian random
variable for small values of $M$. To show this, we set $N_k=2$ and choose
different numbers of transmit antennas, $M$. We generate $10^5$ random
Gaussian distributed channels, $\mathbf{H}_k \in \mathbb{C}^{M \times
N_k}$, and find the left singular matrix of $\mathbf{A}_k \in \mathbb{C}^{M
\times M}$. We randomly pick different entries of $\mathbf{A}_k$. After
normalizing the samples using Lemma 1, we find the 2 bit codebook of
the collected samples using k-means clustering~\cite{Linde}. In
Table~\ref{tab:HaarMatrix} we compare the codebook with that of a unit
variance 2-bit standard Gaussian quantizer~\cite{Lloyd}.
\begin{table}
\caption{Codebook of scalar entries of left singular matrix}
\label{tab:HaarMatrix}
\begin{center}
\begin{tabular}{|l|l|l|l|c|} \hline
M = 2 & M = 3 & M = 4 & M = 8 & Standard Gaussian \\ \hline
-1.34 & -1.40 & -1.43 & -1.48 & -1.51 \\ \hline
-0.43 & -0.44 & -0.45 & -0.45 & 0.45 \\ \hline
0.43 & 0.44 & 0.45 & 0.45 & 0.45 \\ \hline
1.34 & 1.40 & 1.43 & 1.48 & 1.51  \\ \hline
\end{tabular}
\end{center}
\end{table}
In Table~\ref{tab:HaarMatrix}, ``$M = 2$'' stands for the 4 level normalized codebook, 
based on the scalar entries of the
left singular matrix with $2$ transmit antennas. The table shows that even for
small number of transmit antennas (e.g., 3), the probability
distribution of the normalized scalar entries of the Haar matrix resembles the
Gaussian distribution. 

The degrees of freedom of the Haar unitary matrix is less than the total number of real and
imaginary entries. The minimum number of parameters to represent the
Haar matrix can be extracted through Givens' rotations~\cite{Rao}.
An adaptive controller for tracking Givens'
rotated parameters have only been provided for pedestrian
velocities~\cite{Rao}.
The phases and Givens' rotated angles are not Gaussian distributed and least squares based
predictors are not optimum
to track these parameters. Therefore, we stick to our proposed adaptive
differential quantization policy. 
This ensures greater adaptability of our model at the cost 
of slightly higher feedback overhead.

\subsection{Adaptive Differential Quantization of singular value}

The distribution of the square of the singular values of a 
Gaussian channel i.e., the eigenvalues of the Wishart matrix
can be found in~\cite{Karasawa}. We perform fixed 2 bit quantization
of the singular values using this distribution and the standard
Lloyd-Max quantizer. 
%
                                                                    \label{eq:dist_eigval}
%

In this approach, receivers feed back $\mathbf{\hat{F}} =
\mathbf{\hat{H}} \mathbf{\hat{V}}$ to the BS, instead of providing
$\mathbf{\hat{H}}$. The dimensionality of $\mathbf{H}$ and $\mathbf{F}$
are $M \times N$ and $M \times L$ respectively. 
Thus, singular vector quantization saves feedback overhead as long as $L \leq N$.

\subsection{Discussion}
Sections~\ref{sec:DiffQuantChannel} and~\ref{sec:DiffQuantSingVec} show
that, since we assumed the channel to be Gaussian and the difference of
two correlated Gaussian random variables leads to another Gaussian
random variable, the model shown in Fig.~\ref{fig:ADPCM} provides great
flexibility and can hold for different vehicle speeds. To the best of
our knowledge, this is the only work in adaptive differential limited
feedback literature, which can provide both the following advantages:
\begin{enumerate}
\item Unlike the Gauss-Markov models
    of~\cite{Heath:b,Love:b,Love:c}, our model works when the
    normalized autocorrelation between successive channel samples
    drops below 0.5.

\item Unlike the feedback model proposed by~\cite{Rao, Adam:a}, the
    controlling parameters of the predictor and variance estimator
    in our model do not depend on the knowledge of the correlation
    between two successive channel samples.
\end{enumerate}

\section{Numerical Results} \label{sec:NumericalResults}
%

%
\begin{figure}[t]
\begin{center}
 \epsfig{figure=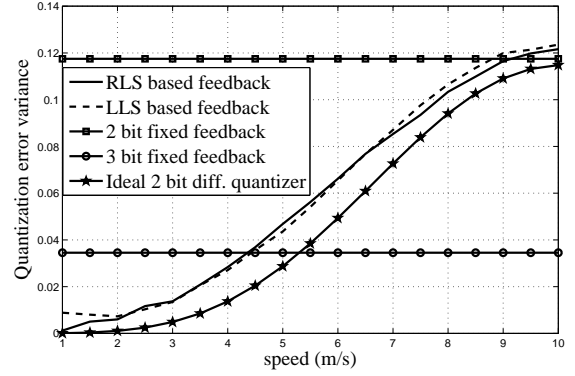,width=85mm}
 \end{center}
 \caption{Comparison of of differential feedback with fixed feedback}  \label{fig:comp_RLS_LLS}
\end{figure} %
The channels were generated using the channel model of~\cite{Zheng} for
each speed. Figure~\ref{fig:comp_RLS_LLS} shows the quantization 
error variance ($\sigma^2_E = E \left[ |h - \hat{h}|^2 \right]$) 
of different proposed methods. The quantization error
variance plot of a 2-bit and 3-bit unit variance fixed Gaussian
quantizer were plotted using the standard values (0.1175 and 0.0345
respectively)~\cite{Lloyd}. The figure shows that the RLS and LLS predictors 
perform very similarly in terms of quantization error reduction. 
The performance of the ideal differential quantizer and predictor is governed by
\begin{eqnarray}
\sigma^2_{d_n} & = & \sigma^2_{h_n} - \psi^H \mathbf{\Phi}^{-1} \mathbf{\psi}  \label{eq:weiner1} \\
\sigma^2_{qn_n} & = & 0.1175 \sigma^2_{d_n} \label{eq:weiner2}
\end{eqnarray}
Here, \eqref{eq:weiner1} follows the minimum error surface of an ideal
Weiner filter~\cite{Haykin} and~\eqref{eq:weiner2} follows the
quantization error associated with a 2-bit, unit-variance, Gaussian
quantizer~\cite{Lloyd}. 
%
%
The RLS adaptive differential
quantizer's performance becomes inferior as the vehicle velocity
exceeds 32 km/h. Using the parameters from
Table~\ref{tab:ChannelParam}, this speed corresponds to a maximum
normalized correlation of 0.0255 between two successive channel
samples. Therefore, our proposed adaptive differential feedback
outperforms fixed quantization as long as the
normalized channel autocorrelation remains positive.
\begin{figure}[t]
\begin{center}
 \epsfig{figure=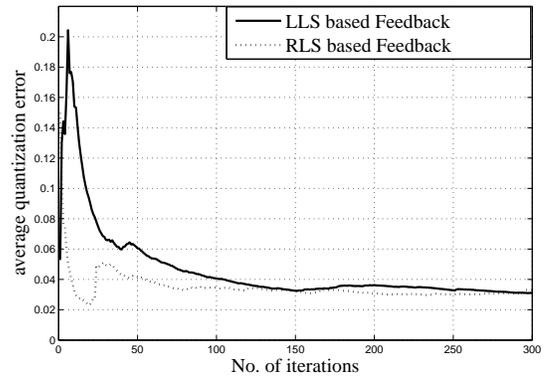,width=77mm}
 \end{center}
 \caption{Comparison of the transient time of RLS and LLS based feedback at 21.6 km/hr}  \label{fig:transient_time}
\end{figure} %

Fig.~\ref{fig:transient_time} shows that the transient time of RLS adaptive differential 
feedback is much smaller than the LLS one. 
The simulation was performed at 21.6 km/hr. The average quantization
error was calculated at every iteration. The
error variance of the proposed quantizer converges close to
its final value within 20 iterations, i.e., 100 ms. 
Thus, the proposed differential
quantizer can adapt itself in real time with reasonable changes in
the mobile velocity. Previous works in differential feedback
literature have either focused on stationary channels with fixed mobile 
velocity~\cite{Love:b,Heath:b,Love:c} or non-stationary channels with
pedestrian velocity~\cite{Rao,Adam:a}. 
Our proposed model are suitable for vehicles whose velocity can increase up to 30 km/hr.

\begin{figure}[t]
\centering
 \epsfig{figure=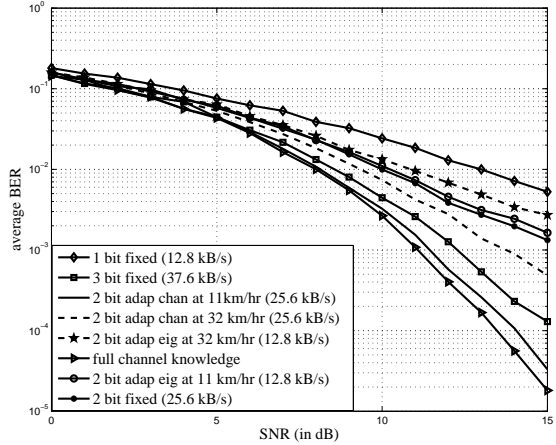,width=85mm}
 \caption{BER and overhead of different feedback methods, $M = 4, K = 2, N_1 = 4, N_2 = 4, L_1 = 2, L_2 = 2$}  \label{fig:BER}
\end{figure}

Figure~\ref{fig:BER} shows the average bit error rate (BER)
performance of different feedback models at different speeds.
We also represent the respective feedback overheads in terms of kilo bit per second (kB/s).
We used the system model of Section II, quadrature phase shift keying modulation and the linear
transceiver design algorithms of~\cite{Khachan} and~\cite{Nazmul:b} to simulate the
performance of channel and singular-matrix quantization.
Here, ``adap chan" and ``adap eig" denote channel quantization and singular vector quantization
respectively. Figure~\ref{fig:BER} shows that the 2-bit adaptive channel entry
feedback outperforms 3-bit fixed feedback and performs very close to
the full channel knowledge feedback scenario at 11 km/h . Even at a high speed of 30 km/h
(corresponds to a normalized autocorrelation of 0.1 with a $0$ degree
arrival angle~\cite{Goldsmith}), the proposed adaptive
feedback reduces the BER by a factor of 2, with respect to 2-bit fixed
feedback per channel entry.

In Fig.~\ref{fig:BER}, ``2 bit adap eig" indicates use of 2 bits to
quantize each of the real and imaginary parts of 
$\mathbf{A}_k$ in an adaptive differential manner. Since, we assumed
$N = 8$ and $L = 4$ in our simulation, spending 2 bits per
scalar singular matrix entry is equivalent to spending 1 bit per real and
imaginary scalar component of the channel. At low
speeds like 11 km/h, the singular matrix entry quantizer performs approximately
as well as the 2-bit fixed quantizer and reduces the feedback overhead
by a factor of 2 for almost same BER. Thus both the adaptive
differential feedback methods save 1 bit per real and imaginary entry
of the channel matrix at low speed (20 km/h for the channel tracker and
8-9 km/h for the singular matrix tracker). This leads to a saving of $2 M N F_s$
bits in feedback overhead per second. Using
Table~\ref{tab:ChannelParam} parameters, the proposed systems provide a feedback
reduction of $12.8$ kBit/sec.

\section{Conclusions} \label{sec:Conclusions}

In this paper, we developed adaptive differential scalar quantization
based limited feedback in a time varying multiuser MIMO channel. The key
contribution is the development of a differential feedback system that
tracks the channel variations without a priori knowledge of the
correlation across time, especially the speed of the vehicle.

We proposed two methods of adaptive differential feedback. First, we
developed 2-bit adaptive differential quantization of each scalar real
and imaginary entry of the MIMO channel. Second, we developed 2-bit
adaptive differential
quantization of each scalar real and imaginary entry of the
singular vectors. Both these methods were shown to significantly
reduce the feedback overhead - by as much as 12.8kBit/sec. Our proposed
adaptive differential quantizer model was shown to be flexible enough
to adapt to different speed of the vehicles.

It is worth emphasizing that one issue not addressed here is adaptive
tracking of the channel gains. When different vehicles are located at
different distances from the base station, we believe that an adaptive
tracking of the channel gain would outperform a fixed channel gain
quantizer.

\bibliographystyle{IEEEbib}
\bibliography{bib_PIMRC_2011}

\end{document}